\makeatletter
\@ifundefined{@parse@version@dash}{%
\def\@parse@version#1{\@parse@version@0#1}
\def\@parse@version@#1/#2/#3#4#5\@nil{%
\@parse@version@dash#1-#2-#3#4\@nil}
\def\@parse@version@dash#1-#2-#3#4#5\@nil{%
  \if\relax#2\relax\else#1\fi#2#3#4 }
}{}
\makeatother

\documentclass[floatfix,aps,prr,superscriptaddress,reprint]{revtex4-2}

\usepackage{bm}
\usepackage{amsmath}
\usepackage{dcolumn}
\usepackage{graphicx}
\usepackage{here}

\begin{document}
\title{Engineering quantum wave-packet dispersion with a strong nonresonant femtosecond laser pulse}
\author{Hiroyuki Katsuki}
\email{katsuki@ms.naist.jp}
\affiliation{Graduate School of Science and Technology, Nara Institute of Science and Technology (NAIST), 8916-5 Takayama-cho, Ikoma, Nara 630-0192, Japan}
\affiliation{Institute for Molecular Science, National Institutes of Natural Sciences, Okazaki 444-8585, Japan}

\author{Yukiyoshi Ohtsuki}
\email{yukiyoshi.ohtsuki.d2@tohoku.ac.jp}
\author{Toru Ajiki}
\affiliation{Department of Chemistry, Graduate School of Science, Tohoku University, 6-3 Aramaki Aza-Aoba, Aoba-ku, Sendai 980-8578, Japan}

\author{Haruka Goto}
\author{Kenji Ohmori}
\email{ohmori@ims.ac.jp}
\affiliation{Institute for Molecular Science, National Institutes of Natural Sciences, Okazaki 444-8585, Japan}
\affiliation{SOKENDAI (The Graduate University for Advanced Studies), Okazaki 444-8585, Japan}

\date{\today}

\begin{abstract}
A non-dispersing wave packet has been attracting much interest from various scientific and technological viewpoints. However, most quantum systems are accompanied by anharmonicity, so that retardation of quantum wave-packet dispersion is limited to very few examples only under specific conditions and targets. Here we demonstrate a conceptually new and universal method to retard or advance the dispersion of a quantum wave packet through “programmable time shift” induced by a strong non-resonant femtosecond laser pulse. A numerical simulation has verified that a train of such retardation pulses stops wave-packet dispersion. 
\end{abstract}

\maketitle

\section{Introduction}
The classical soliton \cite{Scott}, which is a localized wave propagating without spreading, is a general phenomenon that can be observed in various physical systems including water waves \cite{Russell}, optical pulses in a fiber \cite{Mollenauer} and electric LC circuit \cite{Hirota}. In quantum mechanical systems, however, a localized wave packet in general spreads with time due to dispersion induced by anharmonicity. It is a universal phenomenon that is observed in every physical system except for harmonic oscillators, and was discussed in reference to the correspondence between wavefunctions and particles in quantum theory \cite{Schrodinger}. A non-spreading wave packet has been attracting much interest from various scientific and technological viewpoints including pure mathematics \cite{Faddeev, Berry}, many-body physics \cite{Burger, Yefsah, Aycock, Strecker, Khaykovich,Sich, Amo,Egorov} and optical communications \cite{Hasegawa,Spalter}. In very few examples only under specific conditions and targets, however, people could reduce the influence of such anharmonicity by the external perturbation and observe soliton-like motions \cite{Buchleitner}. Those examples include a Bose-Einstein condensate \cite{Burger, Yefsah, Aycock, Strecker, Khaykovich}, Rydberg wave packet in an alkaline atom \cite{Maeda} and microcavity polariton  \cite{Sich, Amo,Egorov}. For example, Maeda {\it et al.} have performed an interesting experiment, in which the Rydberg electron wave packet is irradiated with a microwave continuously, so that the motion of the wave packet is synchronized with the microwave oscillation, and its dispersion is controlled \cite{Maeda}. Although this scheme could be useful for charged particles such as electrons, a more universal scheme is necessary to be applied to electrically neutral systems as well. Here we demonstrate a new experimental method whose concept is universally applicable to the spreading of a wave packet in a variety of quantum mechanical systems. This method utilizes a nonlinear effect induced by a strong non-resonant femtosecond (fs) laser pulse in the near infrared region (NIR pulse). A nonlinear strong-laser effect has been applied to the control of molecular photo-dissociation, population distribution of molecular vibrational levels and selective population of laser-dressed atomic states \cite{Levis, Stolow, Wollenhaupt, Bartels}. Now it is applied to the control of wave-packet spreading. 
Our new method can thus control the shape of a wave packet at any timing during its propagation, and accordingly a sequence of such controls can stop its dispersion, clearly distinguishing itself from previous studies where the free evolution of a wave packet was changed naturally by changing its initial phases \cite{Branderhorst, Kohler, Cao}. 
\begin{center}
\begin{figure}[t]
\includegraphics[width=8.6cm]{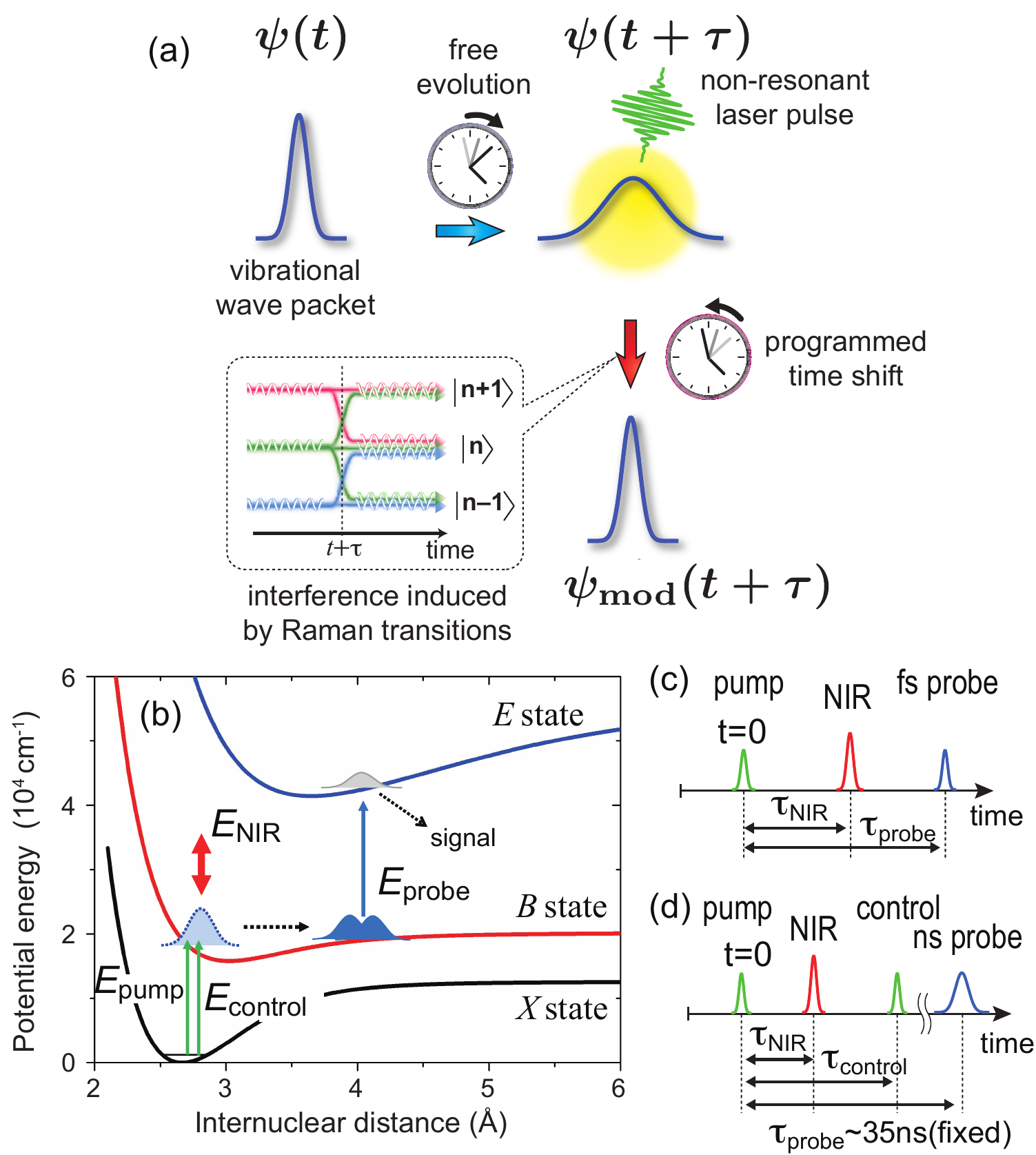}
\caption{Schematic of the spreading-control experiment.
(a) Scenario of the spreading control based on the quantum interference induced by the strong non-resonant NIR pulse. In this sketch, the spreading is retarded by the irradiation of the NIR pulse. 
(b) Potential energy curves of I$_{2}$.  The $B$-state wave packet is created by the pump pulse, and is modulated by the strong NIR pulse at $t=\tau_{\mathrm{NIR}}$. 
The modulated wave packet is excited to the $E$ state by the probe pulse at  $t=\tau_{\mathrm{probe}}$, and its fluorescence signal is detected.
(c) Pulse sequence for the observation of the temporal evolution of the wave packet after the irradiation of the NIR pulse. (d) Pulse sequence for the eigenstate interferogram. }
\end{figure}
\end{center}

\section{Experiment}
A schematic overview of our current experiment is given in Fig. 1 (a).
Our target is a vibrational wave packet generated with a fs pump laser pulse in the $B$ electronic excited state of the I$_{2}$ molecule as shown in Fig. 1(b).
The I$_{2}$ gas is prepared in a vacuum chamber by the  jet expansion of I$_{2}$/Ar mixture through a pinhole (diameter $\sim100$ $\mu$m) at the end of a nozzle.
A vibrational wave packet is created on the $B$ state using the fs pump pulse prepared with an optical parametric amplifier (OPA, Quantronix TOPAS) pumped by a Ti:Sapphire regenerative amplifier (Quantronix TITAN, repetition rate 1kHz, pulse width $\sim$ 100 fs FWHM). The wavelength is set around 535 nm, which is resonant with the $B \gets X (v_{X}=0)$ electronic transition. The bandwidth of the pump pulse is $\sim$ 6 nm, so that the wave packet is composed of $\sim$ 5 eigenstates centered around the vibrational eigenstate $v_{\rm{B}}$=30 of the $B$ state. The pump pulse is thus slightly chirped positively, and its chirp rate is estimated to be $\sim$ 1800 fs$^{2}$.
After the delay $\tau_{\mathrm{NIR}}$, the NIR pulse modulates the wave packet through impulsive Raman transitions \cite{Goto}.
The NIR pulse is prepared with another OPA (Quantronix TOPAS) pumped by the same regenerative amplifier. Its center wavelength is tuned around 1540 nm. The typical power density of the NIR pulse at the sample position is estimated to be $\sim$ 0.8 TW/cm$^{2}$.
At this wavelength, resonant $X \gets B$ transitions induced by the NIR pulse are negligibly small. 
The modulated wave packet is detected as a quantum beat by using another fs probe laser pulse, whose delay is hereafter referred to as $\tau_{\mathrm{probe}}$ (Fig. 1(c)). 
The probe pulse is prepared simultaneously with the NIR pulse using the common OPA. 
Its wavelength is centered around 431 nm. This pulse is resonant with the $E \gets B$ transition of the I$_{2}$ molecule.
The laser induced fluorescence from the $E$ state is collected through a monochromator and with a photo-multiplier.
A quantum beat  is observed as we scan the delay $\tau_{\mathrm{probe}}$.

We also measure the state-resolved time-dependent Ramsey interferogram (hereafter called an eigenstate interferogram) \cite{Katsuki1, Ohmori2, Takei, Liu, Noordam,Bucksbaum1,Scherer1, Leone, Koller,Ohmori,Katsuki2} 
 to obtain the relative phases among different vibrational eigenstates within the wave packet (Fig. 1(d)).
In this Ramsey measurement, we scan the delay $\tau_{\mathrm{control}}$ between the fs pump pulse and its replica (hereafter referred to as a control pulse) with attosecond precision to measure the interferogram of two wave packets produced by a pair of those fs pulses \cite{Ohmori2, Katsuki1, Takei, Liu}.
A  nanosecond (ns) probe pulse around 400 nm is prepared with a dye laser (Lambda Physik Scanmate-2E, Dye: exalite 398, repetition rate 40 Hz). The timing of the ns probe pulse is synchronized with the output of the regenerative amplifier using a frequency divider and a delay generator (SRS DG-535), and is set to $\sim$ 35 ns after the pump pulse. 
The narrow bandwidth of the ns pulse allows for measuring the interferogram of each vibrational eigenstate within the wave packet independently \cite{Katsuki1}.

\section{Experimental results}
Figures 2 (a)--(c) show the quantum beats  measured by scanning the delay $\tau_{\mathrm{probe}}$ of the fs probe pulse as shown in Fig. 1(c). The green trace (a) shows a reference beat measured  without the NIR pulse. The quantum beat with a period of $\sim$ 470 fs is observed, and it corresponds to recurrence motion of the wave packet on the $B$-state potential curve.
The decay of the beat amplitude is due to spreading of the wave packet induced by its dispersion, which arises from the anharmonicity of the electronic potential curve. 
From this green trace, it is seen that the small positive chirp of the pump pulse does not degrade the quantum beat seriously.
The traces (b) and (c) show the beats with the NIR pulse shined at  $\tau_{\mathrm{NIR}}$$\sim$ 5.07 ps (the wave packet  moves around the inner classical turning point) and  $\sim$ 5.36 ps (the wave packet  has just passed the outer classical turning point), respectively.
To remove the effect of the NIR pulse on the wave packet generated in the $E$ state by the probe pulse, the NIR pulse is blocked when the probe pulse is shined before the NIR pulse.
Around the delay $\tau_{\mathrm{probe}}$ $\sim$ 8.5 ps in Fig. 2(a), the wave packet almost collapses and is delocalized, so that the beat amplitude is minimized. 
In Fig. 2 (b), such collapse is advanced and appears immediately after the NIR pulse. In Fig. 2(c), on the other hand, the collapse is retarded toward the opposite direction.
The wave-packet spreading is thus advanced and retarded in Figs. 2(b) and (c), respectively, by the NIR pulses, and their timing difference is only $\sim$ 290 fs, much smaller than the amount of shift of the collapse.
%\onecolumngrid

\begin{center}
\begin{figure*}[t]
\includegraphics[angle=0, width=14.6cm]{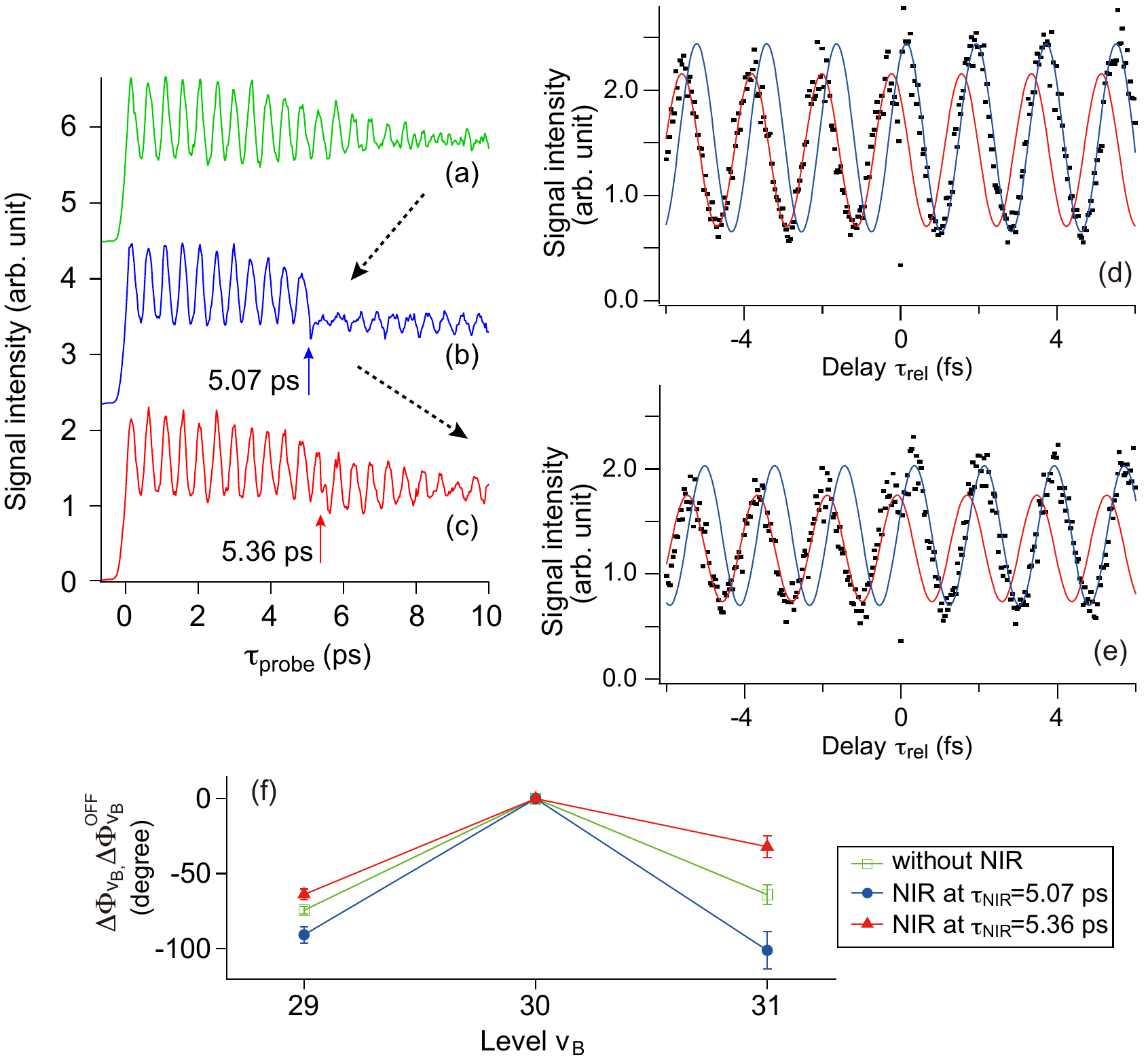}
\caption{
Experimental results of the spreading control.
(a) Free temporal evolution of the wave packet. 
(b) and (c) Actively controlled wave-packet spreading advanced and retarded by the NIR pulse shined at $\tau_{\mathrm{NIR}}$ $\sim$ 5.07 ps and 5.36 ps, respectively.
To remove the effect of the NIR pulse on the wave packet generated in the $E$ state by the probe pulse, the NIR pulse is blocked when the probe pulse is shined before the NIR pulse.
(d) (Dotted line) Eigenstate interferograms of $v_{\rm{B}}$=30 ($\tau_{\mathrm{rel}}<$0) and $v_{\rm{B}}$=29 ($\tau_{\mathrm{rel}}>$0) without the NIR pulse.
(Red line) Fitted sine curve for $v_{\rm{B}}$=30. (Blue line) Fitted sine curve for $v_{\rm{B}}$=29. 
(e) Similar to (d) with the NIR pulse shined at $\tau_{\mathrm{NIR}}$ $\sim$ 5.07 ps.
(f) Relative phases of $v_{\rm{B}}$=29 and 31 to $v_{\rm{B}}$=30 obtained from the eigenstate interferograms exemplified in (d) and (e). The green shows the results without the NIR pulse.  The blue and red show the results with the NIR pulse shined at
$\tau_{\mathrm{NIR}}$ $\sim$ 5.07 ps and $\tau_{\mathrm{NIR}}$ $\sim$ 5.36 ps, respectively, as seen in (b) and (c). }
\end{figure*}
\end{center}

%\twocolumngrid

Next we have measured the state-resolved eigenstate interferograms to see what happens in Figs. 2(b) and (c) to the relative phases among different eigenstates within the wave packet.
Examples of the measured interferograms are plotted with dots in Figs. 2 (d) and (e).
The abscissa is the timing of the control pulse $\tau_{\mathrm{rel}}$=$\tau_{\mathrm{control}} - \tau_{0}$ where $\tau_{0}$ is set around 6.68 ps. In Figs. 2 (d) and (e), the eigenstate observed with the ns probe pulse is switched from the vibrational eigenstate $v_{\rm{B}}$= 30 to $v_{\rm{B}}$= 29 at $\tau_{\mathrm{rel}}$ $\sim$ 0 fs by changing the probe wavelength instantaneously. 
Fig. 2 (d) shows reference interferograms measured without the NIR pulse, whereas Fig. 2 (e) shows the interferogram measured with the NIR pulse shined at $\tau_{\mathrm{NIR}}$ $\sim$5.07 ps. 

Such eigenstate interferograms are measured by scanning  $\tau_{\mathrm{control}}$ around 6.68 ps for the pair of $v_{\rm{B}}$=29 and 30 and the pair of $v_{\rm{B}}$=30 and 31 at the two NIR timings of $\tau_{\mathrm{NIR}}$ $\sim$ 5.07 ps and 5.36 ps corresponding to Figs. 2 (b) and (c), respectively. 
Examples of the eigenstate interferograms measured under different experimental conditions are shown in Fig. S1 in the Supplemental Material \cite{Supp}.
These eigenstate interferograms give the relative phases between $v_{\rm{B}}$=29 and 30 and between $v_{\rm{B}}$=30 and 31 with and without the NIR pulse.
The red and blue solid curves in Figs. 2 (d) and (e) show sine functions 
\begin{equation}
f(\tau_{\mathrm{rel}}) = C+A \sin(\omega \tau_{\mathrm{rel}} + \Phi_{v_{\rm{B}}}) \label{eq1}
\end{equation}
fitted to the measured interferograms in the first half ($\tau_{\rm{rel}} < $0) and the second half ($\tau_{\rm{rel}} > $0), respectively. 
The average frequency of those four fitted sine functions, $\omega_{\mathrm{ave}}$ is assumed to be equal to the average of the transition frequencies of those two vibrational eigenstates, so that $\omega_{\mathrm{ave}}$=($\omega_{30}+\omega_{29}$)/2 or ($\omega_{30}+\omega_{31}$)/2. Using the reported spectroscopic data for $\omega_{29}$, $\omega_{30}$ and $\omega_{31}$ \cite{Luc}, we have calibrated the abscissa of the eigenstate interferograms.  
A set of eigenstate interferograms measured for $v_{\rm{B}}$ = 30 and 29 or $v_{\rm{B}}$ = 30 and 31 with the NIR pulse on or off is then fitted again using the sine functions given as Eq. (\ref{eq1})
 with the fixed transition frequency $\omega_{\mathrm{ave}}$. 
The phase factor $\Phi_{v_{\rm{B}}}$ is hereafter written as $ \Phi_{v_{\rm{B}}} ^{\mathrm{OFF}}$ for the interferogram without the NIR pulse, and as $ \Phi_{v_{\rm{B}}}(\tau_{\rm{NIR}})$ with the NIR pulse shined at the delay $\tau_{\rm{NIR}}$.
We define the relative phase $\Delta \Phi_{v_{\rm{B}}}(\tau_{\mathrm{NIR}})$ of $v_{\rm{B}}$=29  (or 31) to be the phase $\Phi_{v_{\rm{B}}}(\tau_{\mathrm{NIR}})$ of  $v_{\rm{B}}$=29 (or 31) relative to $v_{\rm{B}}$= 30 with the NIR pulse shined at $\tau_{\mathrm{NIR}}$.
Similarly, the relative phase $\Delta \Phi_{v_{\rm{B}}}^{\mathrm{OFF}}$ of $v_{\rm{B}}$=29  (or 31) is defined as the phase of  $v_{\rm{B}}$=29 (or 31) relative to $v_{\rm{B}}$= 30 without the NIR pulse.
 The relation between the phase factor $\Phi_{v_{\rm{B}}}$ and the phase of the eigenfunctions are given in Appendix A.

From the observed interferograms, $\Delta \Phi_{v_{\rm{B}}}(\tau_{\mathrm{NIR}})$ is obtained from each set of eigenstate interferograms measured for $v_{\rm{B}}$ = 30 and 29 or $v_{\rm{B}}$ = 30 and 31 with the NIR pulse on, and averaged over its three independent sets of measurements. 
There are, however, six independent sets of measurements of the phase $\Delta\Phi^{\mathrm{OFF}}_{v_{\rm{B}}}$, three of which combined with the measurements of $\Delta\Phi_{v_{\rm{B}}}$ at $\tau_{\mathrm{NIR}}$ =  5.07 ps and the other three with $\Delta\Phi_{v_{\rm{B}}}$ at $\tau_{\mathrm{NIR}}$ =  5.36 ps, 
 so that we have averaged over those six measurements for $\Delta \Phi ^{\mathrm{OFF}}_{v_{\rm{B}}}$.

In Fig. 2(f) we have plotted $\Delta \Phi_{v_{\rm{B}}}(\tau_{\mathrm{NIR}})$ and  $\Delta \Phi^{\mathrm{OFF}}_{v_{\rm{B}}}$ measured under the three different experimental conditions  corresponding to those shown in Figs. 2(a)--(c).
The shifts of the relative phases $\theta_{v_{\rm{B}}}(\tau_{\mathrm{NIR}})$ induced by the NIR pulse are defined as
\begin{equation}
\theta_{v_{\rm{B}}}(\tau_{\mathrm{NIR}}) = \Delta \Phi_{v_{\rm{B}}}(\tau_{\mathrm{NIR}})-\Delta \Phi^{\mathrm{OFF}}_{v_{\rm{B}}},
\end{equation}
which are  obtained to be $\theta_{29}(\tau_{\mathrm{NIR}}$=5.07 ps)=$-16^{\circ}$, $\theta_{31}(\tau_{\mathrm{NIR}}$=5.07 ps)=$-37^{\circ}$, $\theta_{29}(\tau_{\mathrm{NIR}}$=5.36 ps)=11$^{\circ}$ and $\theta_{31}(\tau_{\mathrm{NIR}}$=5.36 ps)= 32$^{\circ}$, respectively.
It is important to note that the relative phases $\Delta \Phi_{v_{\rm{B}}}$ and $\Delta \Phi ^{\mathrm{OFF}}_{v_{\rm{B}}}$ depend on the choice of $\tau_{0}$, whereas the NIR-pulse induced shift $\theta_{v_{\rm{B}}}$ does not depend on the choice of $\tau_{0}$.
It is thus demonstrated that the relative phases among the eigenstates within the wave packet have been controlled by the NIR pulse, and this phase control leads to the spreading control shown in Figs. 2(b) and (c).

\section{Theoretical analysis and Discussion}
In numerical simulations, we assume the three-electronic-state model as illustrated in Fig. 1(b). 
 The time evolution of the vibrational wave packets in the $E$, $B$ and $X$ states, $|\psi_{E}(t) \rangle$, $|\psi_{B}(t) \rangle$ and $|\psi_{X}(t) \rangle$, is described by the Schr\"{o}dinger equation
 \begin{widetext}
\begin{eqnarray}
& &i\hbar \frac{d}{dt} \left[ 
\begin{array} {l}
 |\psi_{E} (t)\rangle \\
 |\psi_{B} (t)\rangle \\
 |\psi_{X} (t)\rangle 
 \end{array} \right] = \nonumber \\
& &\left[
\begin{array} {ccc}
H_{E}^{0}& -\mu_{EB}(r) E_{\mathrm{probe}}(t-\tau_{\mathrm{probe}}) & 0 \\ 
-\mu_{BE}(r) E_{\mathrm{probe}}(t-\tau_{\mathrm{probe}}) &H_{B}(t) & -\mu_{BX}(r) E(t)  \\ 
0& -\mu_{XB}(r) E(t) & H_{X}(t)
\end{array}
\right]
\left[
\begin{array} {c}
|\psi_{E}(t) \rangle \\ 
|\psi_{B}(t) \rangle   \\ 
|\psi_{X}(t) \rangle 
\end{array}
\right]
. \label{matrix} \nonumber \\
& &
\end{eqnarray}
\end{widetext}
At the initial time, $t_{0}<0$, the molecule is assumed to be in the lowest vibrational state in the $X$ state, $ |\psi_{X} (t_{0})\rangle$=$ |0_{X}\rangle$, with the energy eigenvalue, $\hbar \omega_{0_{X}} \equiv 0$.  Note that we numerically checked that the contributions from the thermally populated vibrational excited states are so small that they do not change the signals in the present study.  
We also note that the rotational period of I$_{2}$ in the $B$ state is $\sim$ 600 ps, 
so that the rotational motion does not affect the signals within the timescale $<$ 10 ps.
The vibrational Hamiltonian of each electronic state is expressed as
 \begin{equation}
H_{e}(t)=H_{e}^{0} -\frac{1}{2} \alpha_{e}(r) [E_{\mathrm{NIR}} (t-\tau_{\mathrm{NIR}})]^{2} \ \ \ \ \ (e=X,B, E), \label{int_H}
\end{equation}
 where $H_{e}^{0}$ and $\alpha_{e}(r)$  are the field-free Hamiltonian and the polarizability function, respectively.
As the $E$ state is solely used to detect the signals after the NIR pulse excitation, we ignore the polarizability interaction in the $E$ state. 
The vibrational eigenstates of $H_{B}^{0}$ and $H_{E}^{0}$ are defined by the eigenvalue problems, $H_{B}^{0}|v_{\rm{B}}\rangle$ = $\hbar \omega_{v_{\rm{B}}} |v_{\rm{B}}\rangle$ and $H_{E}^{0}|v_{E}\rangle$ = $\hbar \omega_{v_{E}} |v_{E}\rangle$ with the vibrational quantum numbers $v_{\rm{B}}$ and $v_{E}$, respectively. The molecular parameters associate with the $B$ and $X$ states are taken from several references which are summarized in our previous study \cite{Ohtsuki}. The $E$-state potential is approximated by the Morse potential, which is generated by using the Rydberg-Klein-Rees (RKR) potential \cite{Barrow}.  The transition moment function between the $E$ and $B$ states is given in Ref. \cite{Akopyan}.

The non-resonant NIR pulse is specified by $E_{\mathrm{NIR}}(t-\tau_{\mathrm{NIR}})$ with $\tau_{\mathrm{NIR}}$ being the time delay with respect to the pump pulse.  Here, we assume that the NIR pulse does not induce the electronic transitions because of its far off-resonant central frequency.  The electronic transitions between the $B$ and $X$ states and between the $E$ and $B$ states are induced by the laser pulse $E(t)$ and the probe pulse $E_{\mathrm{probe}}(t-\tau_{\mathrm{probe}})$, respectively, with the transition moment functions $\mu_{BX}(r)=\mu_{XB}^{\dagger}(r)$  and $\mu_{EB}(r)=\mu_{BE}^{\dagger}(r)$. When simulating the quantum beat observed by a pump-probe scheme, we use  $E(t)$=$E_{\mathrm{pump}}(t)$ in Eq. (\ref{matrix}). The temporal peak of $E_{\mathrm{pump}}(t)$ is set to $t$=0.
 We regard the total population in the $E$ state as the quantum-beat signal, which is expressed as a function of the time delay of the probe pulse, $\tau_{\mathrm{probe}}$. When simulating  the eigenstate interferogram, we use $E(t)$=$E_{\mathrm{pump}}(t)$+$E_{\mathrm{control}}(t-\tau_{\mathrm{control}})$. The control pulse, $E_{\mathrm{control}}(t-\tau_{\mathrm{control}})$, is the replica of the pump pulse but appears with the time delay, $\tau_{\mathrm{control}}$.  The interferogram associated with the $|v_{\mathrm{B}}\rangle$  state is assumed to be proportional to the population of $|v_{\mathrm{B}}\rangle$ after the control pulse. 
 
 We numerically integrate Eq. (\ref{matrix}) by combining the second-order split operator method and fast Fourier transform (FFT).  We adopt the spatial range [2.1 \AA, 6.0 \AA ], which is equally divided into 512 grid points. To reduce the number of the temporal grid points, we assume the rotating-wave approximations for the electronic transitions and introduce the field-interaction representation \cite{Berman}, i.e., a frame rotating at a suitable frequency.  We assume the cycle-averaged polarizability interactions because of the off-resonant frequency of the NIR pulse.
The pump, probe and NIR pulses are assumed to be 80 fs FWHM Gaussian pulses.  We introduce a phase modulation into the pump pulse in the same manner as we did in Ref. \cite{Ohtsuki} to reproduce the experimentally measured quantum beat in Fig. 2(a).  The intensities of the pump and probe pulses are chosen such that they induce less than 5 \% population transitions.  
To reproduce the NIR-pulse effects in Figs. 2(b) and (c), we adjust the intensity of the NIR pulse as well as the polarizability function, $\alpha_{B}(r)$, in the $B$ state.  Roughly speaking, the magnitude of the polarizability interaction is about 4 times larger than that used in our previous study \cite{Ohtsuki}.
Under these conditions, we simulate the signals in Fig. 3(a) in which the timing of the NIR pulse ($\tau_{\mathrm{NIR}}$) is scanned from 5.02 to 5.46 ps. On the other hand, the interferogram of each eigenstate is simulated in the range of $\tau_{\mathrm{control}} \in$ [6.500 ps, 6.505 ps] with 400 signal points.
The number of the signal points is increased by 10 times by means of spline interpolation. At each $\tau_{\mathrm{NIR}}$, we calculate the relative phase by focusing on one of the peaks of the interferogram without the NIR pulse and that with the NIR pulse. 
We repeat the procedure to obtain the relative phases as a function of $\tau_{\mathrm{NIR}}$. By using the relative phases, we calculate the shifts of the relative phases, $\theta_{v_{\rm{B}}}(\tau_{\mathrm{NIR}})$ of $v_{\rm{B}}$=29 and 31 as a function of $\tau_{\mathrm{NIR}}$ as shown in Fig. 3(b).

Figure 3(a) shows that the beat structure is modulated by the NIR pulse, depending on its timing $\tau_{\mathrm{NIR}}$.
Around $\tau_{\mathrm{NIR}}$$\sim$ 5.10 ps, the beat amplitude is dumped immediately after the NIR pulse and is recovered gradually, similar to Fig. 2(b). Around $\tau_{\mathrm{NIR}}$$\sim$ 5.34 ps, the beat amplitude is slightly enhanced by the NIR pulse, and the beat structure lasts  longer  than the one without the NIR pulse, similar to Fig. 2(c).  These NIR timings are similar between the experiments and simulations. 
It is also seen from the comparison between Fig. 2(f) and Fig. 3(b) that $\theta_{v_B}(\tau_{\mathrm{NIR}})$ $<$ 0 at $\tau_{\mathrm{NIR}}$ $\sim$ 5.10 ps and $\theta_{v_B}(\tau_{\mathrm{NIR}})$ $\geq$ 0 at $\tau_{\mathrm{NIR}}$ $\sim$ 5.35 ps for both $v_{\rm{B}}$ = 29 and 31.
The simulations thus qualitatively reproduce the experimental observations, demonstrating that the relative phases among the eigenstates are controlled by the NIR pulse, and this phase control leads to the control of wave-packet spreading.

\begin{center}
\begin{figure}
\includegraphics[angle=0,width=8.5cm]{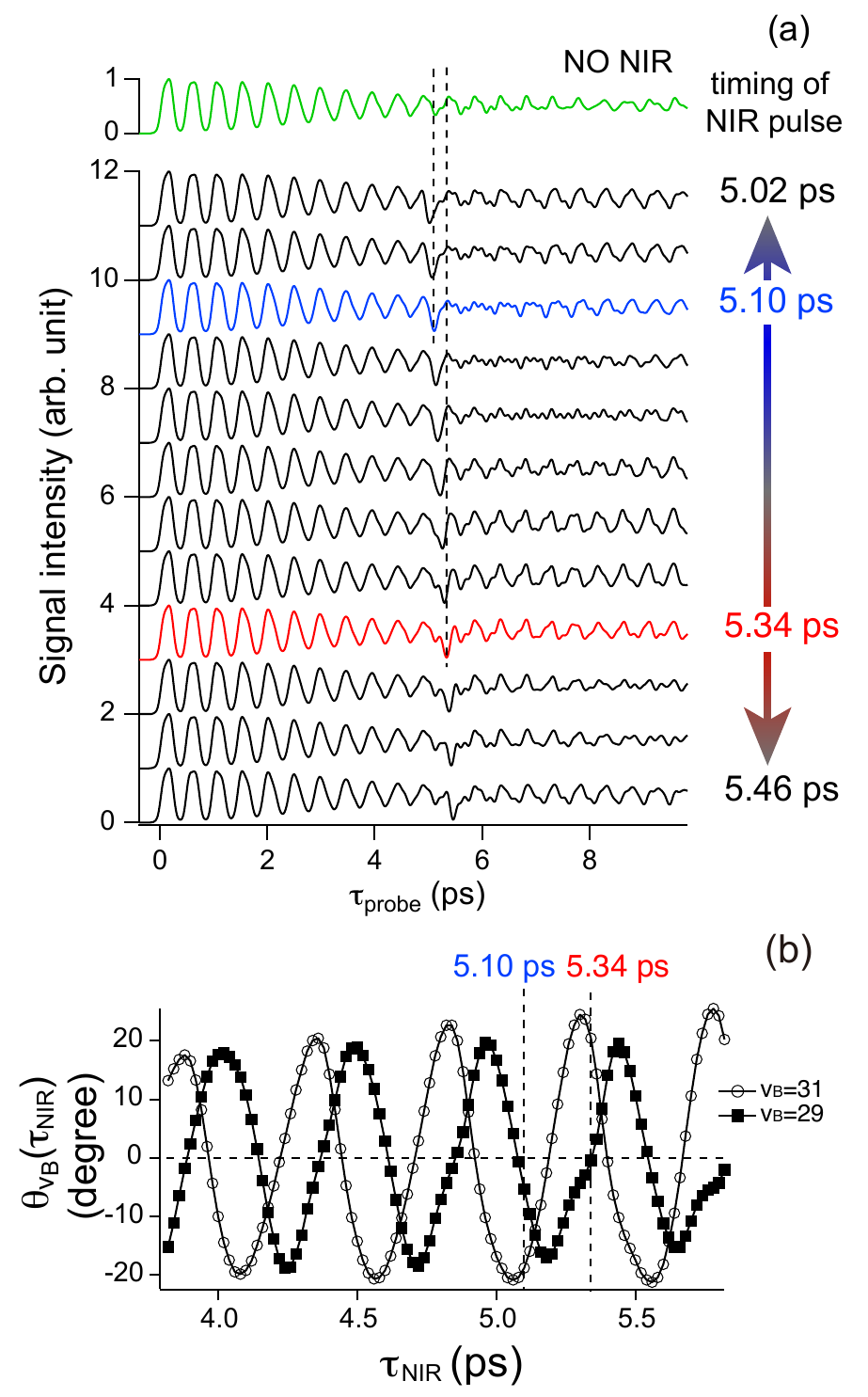}
\caption{(a) Numerical simulation of the quantum beats when the NIR pulse is shined at $\tau_{\mathrm{NIR}}$=5.02 ps to 5.46 ps at every 0.04 ps. The quantum beat at the top (green) is a reference curve simulated without the NIR pulse. The blue trace corresponds to $\tau_{\mathrm{NIR}}$=5.10 ps, in which the NIR pulse is shined at the local minimum after the eleventh beat structure.
The red trace corresponds to $\tau_{\mathrm{NIR}}$=5.34 ps, in which the NIR pulse is shined around a half vibrational period after the blue trace. These two timings are compared to the experimental traces of Fig.2 (b) and (c), where NIR pulses are shined at the similar timings of the traces with the same colors.
(b) Simulated shift of the relative phase $\theta_{v_{\rm{B}}}(\tau_{\mathrm{NIR}})$ of each eigenstate obtained by changing $\tau_{\mathrm{control}}$ from 6.50 ps to 6.50 ps + 5 fs. The solid lines with square and circle markers correspond to $v_{\rm{B}}$=29 and  $v_{\rm{B}}$=31, respectively. The two vertical dotted lines indicate the timings of the NIR pulse $\tau_{\mathrm{NIR}}$=5.10 ps and 5.34 ps.}
\end{figure}
\end{center}

The non-resonant strong NIR pulse induces the Rayleigh and Raman scatterings in the wave packet\cite{Goto}.
We expect that these scattering processes will generate the characteristic coherent mixture depending on the relative phases among the eigenstates involved in the wave packet, which could explain the mechanism underlying the phase control.
For the sake of a qualitative explanation, we derive the analytical expression of the $B$ state wave packet $|\psi_{B}(t)\rangle$  by assuming the first-order approximations with respect to the pump-pulse excitation and the Raman scattering.  
The matrix element associated with the Raman scattering ($v_{\rm{B}} \neq v_{\rm{B}}'$) is given by
\begin{multline}
\hbar \gamma(v_{\rm{B}},v_{\rm{B}}') = \\
\langle v_{\rm{B}} | \alpha_{B}(r)|v_{\rm{B}}'\rangle \int_{-\infty}^{\infty} ds e^{-i(\omega_{v_{\rm{B}}}-\omega_{v'_{\rm{B}}})s}[E_{\mathrm{NIR}}(s)]^{2}, \label{E_shift}
\end{multline}
where the value is assumed to be real.  
Here,  $\alpha_{B}(r)$ is the polarizability function of the $B$ state, and  $\hbar\omega_{v_{\rm{B}}}$ ($\hbar\omega_{v'_{\rm{B}}}$) is the energy eigenvalue of the vibrational eigenstate  $|v_{\rm{B}}\rangle$ ($|v'_{\rm{B}}\rangle$).  Equation (\ref{E_shift}) shows that $\gamma(v_{\rm{B}},v'_{\rm{B}})$ is expressed as the matrix element $\langle v_{\rm{B}}|\alpha_{B}(r)|v'_{\rm{B}}\rangle$ multiplied by the frequency component of $[E_{\mathrm{NIR}}(t)]^{2}$ corresponding to the $v_{\rm{B}}-v'_{\rm{B}}$ Raman transition.
We further assume that the $|\Delta v_{\rm{B}}|=1$ Raman transitions dominate the $|\Delta v_{\rm{B}}| \ge 2$ transitions and neglect the quantum number dependence, $\gamma(v_{\rm{B}},v_{\rm{B}}\pm1) \simeq -|\gamma|$,  the negative value of which is assumed according to our previous study\cite{Ohtsuki}.
We then have the expression of the probability amplitude as
\begin{eqnarray}
\langle v_{\rm{B}}| \psi_{B}(t)\rangle &=&e^{-i\omega_{v_{\rm{B}}}t+i\eta}\left[ \left\langle v_{\rm{B}}|\psi_{B}^{0_{X}}\right \rangle \right.  \nonumber \\
&&-  \frac{i}{2}|\gamma| \left\{ e^{-i(\omega_{v_{\rm{B}}+1}-\omega_{v_{\rm{B}}})\tau_{\mathrm{NIR}}} \langle v_{\rm{B}}+1|\psi_{B}^{0_{X}} \rangle \right. \nonumber \\
&&+ \left. \left. e^{i(\omega_{v_{\rm{B}}}-\omega_{v_{\rm{B}}-1})\tau_{\mathrm{NIR}}} \langle v_{\rm{B}}-1|\psi_{B}^{0_{X}} \rangle \right \} \right ], \label{int_H2}
\end{eqnarray}
where we have assumed the quantum-number independent shift, $\eta \simeq\gamma(v_{\rm{B}},v_{\rm{B}})/2$.
In Eq. (\ref{int_H2}), the wave packet initially excited by the pump pulse, $|\psi_{B}^{0_{X}}\rangle$, is expressed as
\begin{equation}
|\psi_{B}^{0_{X}}\rangle=\int_{-\infty}^{\infty} dt_{1} e^{i H_{B}^{0} t_{1}/{\hbar}} E_{\mathrm{pump}}(t_{1})\mu_{BX}(r)|0_{X} \rangle \label{eq7}
\end{equation}
in the first-order perturbation approximation.  The derivation is summarized in Appendix B.
Briefly, the first line of Eq. (\ref{int_H2}) describes the component whose $v_{\rm{B}}$  is unchanged by the NIR pulse, and the second and third lines represent the contributions from the neighboring  $|v_{\rm{B}}\pm1\rangle$ states through the Raman transitions, respectively.  Equation (\ref{int_H2}) thus shows that the neighboring states, $|v_{\rm{B}}\pm1\rangle$, are coherently mixed into the state $|v_{\rm{B}}\rangle$  through the Raman transitions induced by the NIR pulse, and this mixing leads to the experimentally observed phase shifts.  It is important to note that this control scenario is based on the Raman transitions, which is not specific to molecular eigenstates, but universal to any type of Raman-active eigenstates of a variety of quantum systems \cite{Harrison, Tosi, Wertz}.

From the numerical simulation in Fig. 3(b), we note that there appear interesting timings when
  $\theta_{29}(\tau_{\mathrm{NIR}})$=$-\theta_{31}(\tau_{\mathrm{NIR}})>$ 0 for every vibrational period. 
At these timings, we can undo the wave-packet spreading.
The NIR pulse shined at those timings to reverse the wave-packet spreading is hereafter referred to as an ``undo pulse."
For example, $\tau_{\mathrm{NIR}}$$\sim$ 5.48 ps gives the phase shift $\theta_{29}(\tau_{\mathrm{NIR}}) \simeq -\theta_{31}(\tau_{\mathrm{NIR}}) \sim 15^{\circ}$, shifting the time backward by 475 $\times$ 15/360 $\simeq$ 20 fs, where 475 fs is a classical oscillation period of the wave packet. 
It is also possible to shift the time forward by almost the same amount if we choose $\tau_{\mathrm{NIR}}$$\sim$ 5.23 ps when $\theta_{29}(\tau_{\mathrm{NIR}})$=$-$ $\theta_{31}(\tau_{\mathrm{NIR}})<$ 0 holds.
A train of the undo pulses could stop wave-packet spreading, leading to a new concept of dispersion management.

To test this concept, we have numerically designed the undo pulses by means of quantum optimal control theory, which will be discussed elsewhere\cite{Ohtsuki2}.  Here, instead of the full optimization approach, we consider an analytical approach by adopting a minimal dephasing model that consists of three vibrational states in the $B$ state, $|v_{\rm{B}}\rangle$, $|v_{\rm{B}}+1\rangle$, and $|v_{\rm{B}}+2\rangle$ with a specific initial condition, $\langle v_{\rm{B}}|\psi_{B}^{0_{X}}\rangle$ = $\langle v_{\rm{B}}+1|\psi_{B}^{0_{X}}\rangle$ = $\langle v_{\rm{B}}+2|\psi_{B}^{0_{X}}\rangle$.
The frequency differences are given by  $\omega_{v_{\rm{B}}+2}-\omega_{v_{\rm{B}}+1} = \omega - \Delta\omega$ and $\omega_{v_{\rm{B}}+1}-\omega_{v_{\rm{B}}} = \omega + \Delta\omega$ ($\Delta\omega>0$), in which the anharmonic frequency, $\Delta\omega$, causes the wave-packet spreading.
We introduce the vibrational period, $T$=$2\pi$/$\omega$, and the  dimensionless parameter, $\delta=\Delta \omega/\omega$. 
In the following, we examine how to suppress modulation of the oscillating part of the quantum beat by using a NIR-pulse train because the degree of modulation directly reflects the degree of wave-packet spreading.  When deriving the analytical expressions, we adopt the same assumptions and approximations as those we used when deriving Eq. (\ref{int_H2}).  We also neglect the vibrational quantum number dependence of the overlap integrals associated with the $E\gets B$ transitions.
According to Appendix B, the oscillating part of the quantum beat in the absence of the NIR pulse is expressed as a function of a time delay,  $x$=$\tau_{\mathrm{probe}}$/$T$,
\begin{equation}
S_{\mathrm{QB}}^{\mathrm{(osc)}}(x)=-\cos(2\pi\delta x) \cos(2\pi x),	\label{tr_QB}
\end{equation}
except for unimportant factors.
We consider the NIR pulse train composed of $\{ E_{\mathrm{NIR}}^{(n)}(t-\tau_{\mathrm{NIR}}^{(n)} ); n=1,2, \cdots N\}$ 
with $\tau_{\mathrm{NIR}}^{(n)}$ being the time delay of the $n$-th NIR pulse.  We add the suffix $n$ to the matrix element in (\ref{E_shift}), $\gamma^{(n)}(v_{\rm{B}},v_{\rm{B}}')\simeq -|\gamma^{(n)}|$, to specify that it is originated from the $n$-th NIR pulse, $E_{\mathrm{NIR}}^{(n)}(t-\tau_{\mathrm{NIR}}^{(n)} )$.
Then, the oscillating part of the quantum beat induced by the lowest-order Raman transitions can be expressed as
\begin{multline}
S_{\mathrm{NIR}}^{\mathrm{(osc)}}(x)=\\
\frac{1}{2} \sum_{n=1}^{N} |\gamma^{(n)}| \sin \left[ 2\pi\delta(x-y^{(n)}) \right] \cos \left[ 2\pi (x+y^{(n)}) \right], \label{NIR_osc}
\end{multline}
where the dimensionless time delay is defined by $y^{(n)}=\tau_{\mathrm{NIR}}^{(n)} /T$.

\begin{center}
\begin{figure}[t]
\includegraphics[angle=0, width=8.5cm]{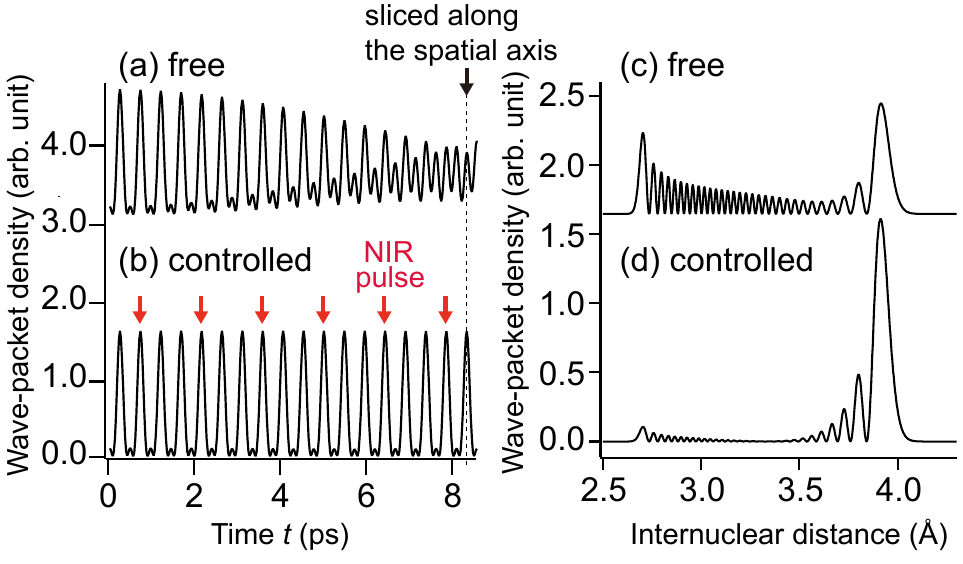}
\caption{Simulated wave-packet spreading suppressed by a train of strong NIR pulses.
The wave packet here is an isolated three-level system composed of the eigenstates $v_{\rm{B}}$ = 29, 30 and 31 of the iodine molecule with their population ratio 1:1:1 at the time origin $t$ = 0. (a) and (b) The wave-packet density at the inter-atomic distance of 3.92 \AA \  is plotted as a function of time {\it t}. (c) and (d) The wave-packet density at {\it t} = 8.315 ps is plotted as a function of the inter-atomic distance.
(a) and (c) Free wave packet without NIR pulses. (b) and (d) Wave packet controlled with a train of six NIR pulses shined at $t$ =$(3n-1.5)T$ with $n$=1, 2, ..., 6 where $T$= 475 fs is a classical oscillation period of the wave packet. These timings are indicated by the arrows in Fig. 4(b). 
Each NIR pulse has a Gaussian envelope with a temporal width of 0.05 $T$ (FWHM). 
The spreading of the wave packet is stopped almost completely by the NIR pulses.  Plots (a) and (c) are vertically offset for better visibility.
The movies of these wave packets are given in Video S1 in the Supplemental Material.}
\end{figure}
\end{center}

As a simple example, we examine how to suppress the dephasing by using a single NIR pulse.  Because the anharmonicity $\delta$ appears in the envelope in Eq. (\ref{tr_QB}), we need to find the right timing and intensity of the NIR pulse to replace $\cos(2\pi \delta x)$ by $\cos [2\pi \delta (x-X^{(1)})],$ where $X^{(1)}>$ 0  specifies the temporal backward shift.  This has to be done by introducing the signal in Eq. (\ref{NIR_osc}) with $N$=1, which leads to the conditions, $X^{(1)}=k^{(1)}$, $y^{(1)}=k^{(1)}/2$, and $|\gamma^{(1)}|=4\sin (\pi\delta k^{(1)})$ with $k^{(1)}=1,3,5,\cdots$.
It means that the NIR pulse should be applied when the wave packet is located at the outer turning point, which is consistent with the experimental observation in Fig. 2(c) and with the simulation in Fig. 3(a).  
It is straightforward to extend the analyses of Eq. (\ref{NIR_osc}) with a general case of $N$ to suppress the modulation in the quantum-beat signal, i.e., to suppress the wave-packet spreading.  Note that it is also straightforward to see that the change of the sign of  $\gamma^{(n)}$ will shift the timing of the NIR pulse by a half of the vibrational period, $T$/2.

This new concept of dispersion management has been demonstrated by numerical simulations in the three-level model system composed of $v_{\rm{B}}$=29, 30 and 31 of the iodine molecule, which leads to the period, $T$=0.475 ps  and the parameter, $\delta$=0.015.  
As an example, we consider the case of $k^{(1)}$, $k^{(2)}$, $\cdots$, $k^{(N)}$=3 with $N$=6. 
Each NIR pulse in the pulse train is assumed to have a Gaussian envelope with a common temporal width of, $2\sqrt{\ln 2}\sigma$=0.05, (FWHM in units of $T$), corresponding to $\sim$ 24 fs.  Note that the time evolution of the controlled wave packet is independent of the choice of the value of  $\sigma$ if it is smaller than 0.1 (in units of $T$).  Figure 4(a) shows a temporal evolution of the wave-packet density at the inter-atomic distance of 3.92 \AA \  without the undo pulses.  The beat amplitude decreases due to the spreading of the wave packet.  Figure 4(b) shows that this spreading is stopped almost completely by a train of the undo pulses shined at the timings indicated by the analytical treatment, and with arrows in the figure.  Accordingly, the shapes are dramatically different at $t$ =8.315 ps between these two wave packets, whose snapshots are shown in Figs. 4(c) and (d).  The movies of these wave packets are given in Video S1 in the Supplementary Material \cite{Supp}.

\section{Conclusion}
We have demonstrated the control of wave-packet dispersion with a strong non-resonant NIR pulse.
The relative phase shifts induced by the NIR pulse among the vibrational eigenstates within the wave packet have been measured by the time-dependent Ramsey interferometry with attosecond precision.
It is shown that we can advance or retard the wave-packet motion by tuning the timing of the NIR pulse.
The model simulation assuming Raman transitions between neighboring vibrational levels reproduces the experimental results faithfully, supporting 
our new concept of dispersion management by strong-laser-induced interference.
Extending this scheme with a train of the NIR pulses, we have demonstrated a non-dispersing wave packet by numerical simulations in the three-level model system. 
 Such a simplest three-level system could actually be prepared in artificial quantum systems including quantum dots and wells \cite{Harrison}. 
This scenario of dispersion management is universal to any Raman-active quantum systems.
Those systems would include a variety of electronic states in Rydberg atoms and quantum materials.

\begin{acknowledgments}
This work is supported by JSPS KAKENHI Grant Numbers JP16H06289 (Specially Promoted Research), JP21245007, JP20K05414 and JP21H01890. This work is also supported by MEXT ``Photon-Frontier-Network" program and by MEXT Quantum Leap Flagship Program (MEXT Q-LEAP) JPMXS0118069021. 
\end{acknowledgments}

\appendix
\section{Relation between the observed signal and phase of eigenfunctions}
We consider the phase factor $e^{-i(\omega_{v_{\rm{B}}} t + \phi_{v_{\rm{B}}})}$ of the vibrational eigenstate $|v_{\rm{B}} \rangle$, where $\omega_{v_{\rm{B}}}$ is its angular frequency and $\phi_{v_{\rm{B}}}$  is the phase factor of the $v_{\rm{B}}$ state. 
The phase $\phi_{v_{\rm{B}}}$ is further decomposed as $\phi_{v_{\rm{B}}}=\phi_{v_{\rm{B}}}^{0}+\delta_{v_{\rm{B}}}(\tau_{\mathrm{NIR}})$ where $\phi_{v_{\rm{B}}}^{0}$ is an initial phase offset determined by the characteristics of the pump pulse, and $\delta_{v_{\rm{B}}}(\tau_{\mathrm{NIR}})$ is a phase shift induced by the NIR pulse. 
Note that the second term does not appear without the NIR pulse.
The superposition of two wave packets generated by the pump and control pulses whose delay is given by  $\tau_{\mathrm{control}}=\tau_{0}+\tau_{\mathrm{rel}}$ can be written as $\sum_{v_{\rm{B}}} e^{-i\omega_{v_{\rm{B}}}t -i\phi_{v_{\rm{B}}}^{0}}|v_{\rm{B}} \rangle (1+e^{i\omega_{v_{\rm{B}}} (\tau_{0}+\tau_{\mathrm{rel}})})$, which gives the interferogram $\propto (1+\cos(\omega_{v_{\rm{B}}} \tau_{\mathrm{rel}}+\phi_{v_{\rm{B}}}^{\mathrm{res}}))$, where $\phi_{v_{\rm{B}}}^{\mathrm{res}}$ is the NIR independent residual phase term. 
When the NIR pulse induces a phase shift $\delta_{v_{\rm{B}}}(\tau_{\mathrm{NIR}})$, the superposition is  written as $\sum_{v_{\rm{B}}} e^{-i\omega_{v_{\rm{B}}}t-i \phi_{v_{\rm{B}}}^{0}-i\delta_{v_{\rm{B}}}} |v_{\rm{B}} \rangle (1+e^{i\omega_{v_{\rm{B}}} (\tau_{0}+\tau_{\mathrm{rel}})+i\delta_{v_{\rm{B}}}})$, which gives the interferogram $\propto (1+\cos(\omega_{v_{\rm{B}}} \tau_{\mathrm{rel}}+ \phi_{v_{\rm{B}}}^{\mathrm{res}}+\delta_{v_{\rm{B}}}))$. 
Comparing the two interferograms, it is seen that the phase shift $\delta_{v_{\rm{B}}}(\tau_{\mathrm{NIR}})$ induced by the NIR pulse directly appears as the phase shift between the eigenstate interferograms with and without the NIR pulse.

The relation between $\theta_{v_{\rm{B}}}(\tau_{\mathrm{NIR}})$ and $\delta_{v_{\rm{B}}}(\tau_{\mathrm{NIR}})$ is given as
\begin{eqnarray}
\theta_{v_{\rm{B}}}(\tau_{\mathrm{NIR}})&=&\Delta \Phi_{v_{\rm{B}}}(\tau_{\mathrm{NIR}}) - \Delta \Phi ^{\mathrm{OFF}}_{v_{\rm{B}}} \nonumber \\
&=&  (\Phi_{v_{\rm{B}}}(\tau_{\mathrm{NIR}})-\Phi_{30}(\tau_{\mathrm{NIR}})) - ( \Phi ^{\mathrm{OFF}}_{v_{\rm{B}}} - \Phi ^{\mathrm{OFF}}_{30})  \nonumber \\
&=& (\Phi_{v_{\rm{B}}}(\tau_{\mathrm{NIR}}) - \Phi ^{\mathrm{OFF}}_{v_{\rm{B}}}) -(\Phi_{30}(\tau_{\mathrm{NIR}}) - \Phi ^{\mathrm{OFF}}_{30}) \nonumber \\
&=&\delta_{v_{\rm{B}}}(\tau_{\mathrm{NIR}}) -\delta_{30}(\tau_{\mathrm{NIR}})		\label{def_ph}
\end{eqnarray}
where  $\delta_{v_{\rm{B}}}(\tau_{\mathrm{NIR}})$ is the phase shift of $v_{\rm{B}}$ = 29, 30 and 31 induced by the NIR pulse defined in the beginning of this appendix.

\section{Analytical expression of the quantum beats}
If we assume the first-order approximation with respect to the pump-pulse excitation, the $B$-state wave packet after the irradiation of the NIR pulse is expressed as
\begin{equation}
|\psi_{B}(t) \rangle  =\frac{i}{\hbar}\int_{-\infty}^{\infty} dt_{1} U_{B}(t,t_{1}) \mu_{BX}(r)E_{\mathrm{pump}}(t_{1}) |0_{X}\rangle \label{B1}
\end{equation}
where the time evolution operator that includes the NIR-pulse-induced interaction is defined by
\begin{equation}
U_{B}(t_{2},t_{1})=\hat{T} \exp \left[ -\frac{i}{\hbar}\int_{t_{1}}^{t_{2}} dsH_{B}(s) \right] \label{ord}
\end{equation}
with the time-ordering operator, $\hat{T}$. 
According to our previous study, we divide the NIR-pulse-induced interaction into two parts,
\begin{multline}
-\frac{1}{2} \alpha_{B}(r) [E_{\mathrm{NIR}} (t-\tau_{\mathrm{NIR}})]^{2} = \\
-\frac{1}{2}\sum_{v_{\rm{B}}} |v_{\rm{B}}\rangle \langle v_{\rm{B}}|\alpha_{B}(r) [E_{\mathrm{NIR}}(t-\tau_{\mathrm{NIR}})]^{2}|v_{\rm{B}}\rangle \langle v_{\rm{B}}| +\bar{V}_{B}^{\alpha}(t), \label{Rm}
\end{multline}
where the first (second) term of the right-hand side mainly induces the Rayleigh scattering (Raman scattering).
That is, the first term in the right-hand side of Eq. (\ref{Rm}) leads to the energy shift of the $|v_{\rm{B}}\rangle$ state, whose vibrational quantum number dependence can be neglected as shown in our previous study \cite{Ohtsuki}.
Because the phase shift, $\gamma(v_{\rm{B}},v_{\rm{B}})$, is virtually independent of the vibrational quantum number, it would be straightforward to rewrite Eq. (\ref{B1}) as 
\begin{equation}
|\psi_{B}(t) \rangle  =\frac{i}{\hbar} e^{-iH_{B}^{0}t/\hbar} U_{B}^{(I)}(\tau_{\mathrm{NIR}}) |\psi_{B}^{0_{X}}\rangle,  \label{B4}
\end{equation}
where $ |\psi_{B}^{0_{X}}\rangle$ is given by Eq. (\ref{eq7}) and $U_{B}^{(I)}(\tau_{\mathrm{NIR}})$ is defined by
\begin{equation}
U_{B}^{(I)}(\tau_{\mathrm{NIR}})=\hat{T} \exp \left[ -\frac{i}{\hbar}\int_{-\infty}^{\infty} ds\ e^{iH_{B}^{0}s/\hbar} \bar{V}_{B}^{\alpha}(s) e^{-iH_{B}^{0}s/\hbar} \right],
\end{equation}
with the operator $\bar{V}_{B}^{\alpha}(s)$ in Eq. (\ref{Rm}).
We expand $U_{B}^{(I)}(\tau_{\mathrm{NIR}})$  up to the first order with respect to $\bar{V}_{B}^{\alpha}(s)$ and introduce the assumptions and approximations as explained above Eq. (\ref{int_H2}).  We then obtain Eq. (\ref{int_H2}).

We next derive the expression of the vibrational quantum-beat signal after the probe pulse by assuming the first-order perturbation approximation with respect to the probe-pulse excitation and by using Eq. (\ref{B4}).  Because the signal is assumed to be proportional to the $E$-state population, the signal is expressed as
\begin{equation}
S_{\mathrm{QB}}(\tau_{\mathrm{probe}})=\left| \left\langle  \psi_{E}^{v_{E}} \left| e^{-iH_{B}^{0}\tau_{\mathrm{probe}}/\hbar}  U_{B}^{(I)}(\tau_{\mathrm{NIR}})  \right| \psi_{B}^{0_{X}}  \right\rangle \right|^{2}, \label{QB_pr}
\end{equation}
except for unimportant factors.
Here, we have assumed a single vibrational state, $|v_{E}\rangle$ (the energy eigenvalue, $\hbar\omega_{v_{E}}$), in the $E$ state for convenience.
In Eq. (\ref{QB_pr}), $\langle\psi_{E}^{v_{E}}|$ is defined by
\begin{equation}
\langle\psi_{B}^{v_{E}}|= \langle v_{E}| \mu_{EB}(r) \int_{-\infty}^{\infty} dt_{2} e^{-i\left(H_{B}^{0}/\hbar-\omega_{v_{E}}\right) t_{2}} E_{\mathrm{probe}}(t_{2}).
\end{equation}
Expanding $U_{B}^{(I)}(\tau_{\mathrm{NIR}})$  in the power of  $\bar{V}_{B}^{\alpha}(s)$, we see that the zeroth order term leads to the signal in the absence of the NIR pulse.  We assume a real probability amplitude, $a_{v_{\rm{B}}}=\langle v_{\rm{B}}|\psi_{B}^{0_{X}} \rangle$, and neglect the vibrational quantum number dependence of the overlap integral, $\langle v_{\rm{B}}|\psi_{B}^{v_{E}} \rangle \langle\psi_{B}^{v_{E}}|v_{\rm{B}}'\rangle$.  If we consider the case where the probe pulse detects the wave packet around the outer turning point and extract the oscillating part of the signal with $\Delta v_{\rm{B}}$=$\pm$1, then we will have
\begin{equation}
S_{\mathrm{QB}}^{(\mathrm{osc})}(\tau_{\mathrm{probe}})=-\sum_{v_{\rm{B}}}a_{v_{\rm{B}}}a_{v_{\rm{B}}-1}\cos[(\omega_{v_{\rm{B}}}-\omega_{v_{\rm{B}}-1})\tau_{\mathrm{probe}}], \label{QB_osc}
\end{equation}
except for unimportant factors.  

When considering the NIR-pulse-induced quantum beat, we assume the pulse train composed of $N$ NIR pulses, which is expressed as $\sum_{n=1}^{N} E_{\mathrm{NIR}}^{(n)}(t-\tau_{\mathrm{NIR}}^{(n)})$, with $\{\tau_{\mathrm{NIR}}^{(n)}\}$ being the time delays.  Because of this modification, the matrix element, $\gamma(v_{\rm{B}},v_{\rm{B}}')$ in Eq. (5), is replaced with $\gamma^{(n)}(v_{\rm{B}},v_{\rm{B}}')$, where the superscript, $(n)$, is introduced to denote that the matrix element is originated from the $n$-th out of $N$ NIR pulses, $E_{\mathrm{NIR}}^{(n)}(t-\tau_{\mathrm{NIR}}^{(n)})$.  We extract the first-order term with respect to $\bar{V}_{B}^{\alpha}(s)$  from Eq. (B6) and adopt the same assumptions and approximations we used in deriving Eqs. (6) and (B8).  Then we obtain the oscillating component of the first-order term,

 \begin{widetext}
 \begin{multline}
S_{\mathrm{NIR}}^{(\mathrm{osc})}(\tau_{\mathrm{probe}})= -\sum_{n=1}^{N} \sum_{v_{\rm{B}}}|\gamma^{(n)}| \left\{ a_{v_{\rm{B}}+1}^{2}\sin \left[(\omega_{v_{\rm{B}}+1}-\omega_{v_{\rm{B}}})(\tau_{\mathrm{probe}}-\tau_{\mathrm{NIR}}^{(n)})\right] \right. \\
\left. - a_{v_{\rm{B}}-1}^{2} \sin \left[(\omega_{v_{\rm{B}}}-\omega_{v_{\rm{B}}-1}) (\tau_{\mathrm{probe}}-\tau_{\mathrm{NIR}}^{(n)})\right] \right\} \\
-\sum_{n=1}^{N} \sum_{v_{\rm{B}}}|\gamma^{(n)}|  a_{v_{\rm{B}}+1}a_{v_{\rm{B}}-1} \left\{  \sin \left[ (\omega_{v_{\rm{B}}+1}-\omega_{v_{\rm{B}}})  \tau_{\mathrm{probe}}+
(\omega_{v_{\rm{B}}}-\omega_{v_{\rm{B}}-1})\tau_{\mathrm{NIR}}^{(n)} \right] \right. \\
\left. -\sin\left[ (\omega_{v_{\rm{B}}}-\omega_{v_{\rm{B}}-1}) \tau_{\mathrm{probe}}+(\omega_{v_{\rm{B}}+1}-\omega_{v_{\rm{B}}}) \tau_{\mathrm{NIR}}^{(n)} \right] \right\} \label{QBsig}
\end{multline}
\end{widetext}
which is applied to the three-state model to derive Eq. (9).

\end{document}

\title{Supplemental Material for ``Engineering quantum wave-packet dispersion with a strong nonresonant femtosecond laser pulse''}

\author{Hiroyuki Katsuki}
\email{katsuki@ms.naist.jp}
\affiliation{Graduate School of Science and Technology, Nara Institute of Science and Technology (NAIST), 8916-5 Takayama-cho, Ikoma, Nara 630-0192, Japan}
\affiliation{Institute for Molecular Science, National Institutes of Natural Sciences, Okazaki 444-8585, Japan}

\author{Yukiyoshi Ohtsuki}
\email{yukiyoshi.ohtsuki.d2@tohoku.ac.jp}
\author{Toru Ajiki}
\affiliation{Department of Chemistry, Graduate School of Science, Tohoku University, 6-3 Aramaki Aza-Aoba, Aoba-ku, Sendai 980-8578, Japan}

\author{Haruka Goto}
\author{Kenji Ohmori}
\email{ohmori@ims.ac.jp}
\affiliation{Institute for Molecular Science, National Institutes of Natural Sciences, Okazaki 444-8585, Japan}
\affiliation{SOKENDAI (The Graduate University for Advanced Studies), Okazaki 444-8585, Japan}

\date{\today}

\maketitle

\newpage
\setcounter{figure}{0}
\begin{figure}[H]
\includegraphics[scale=0.7]{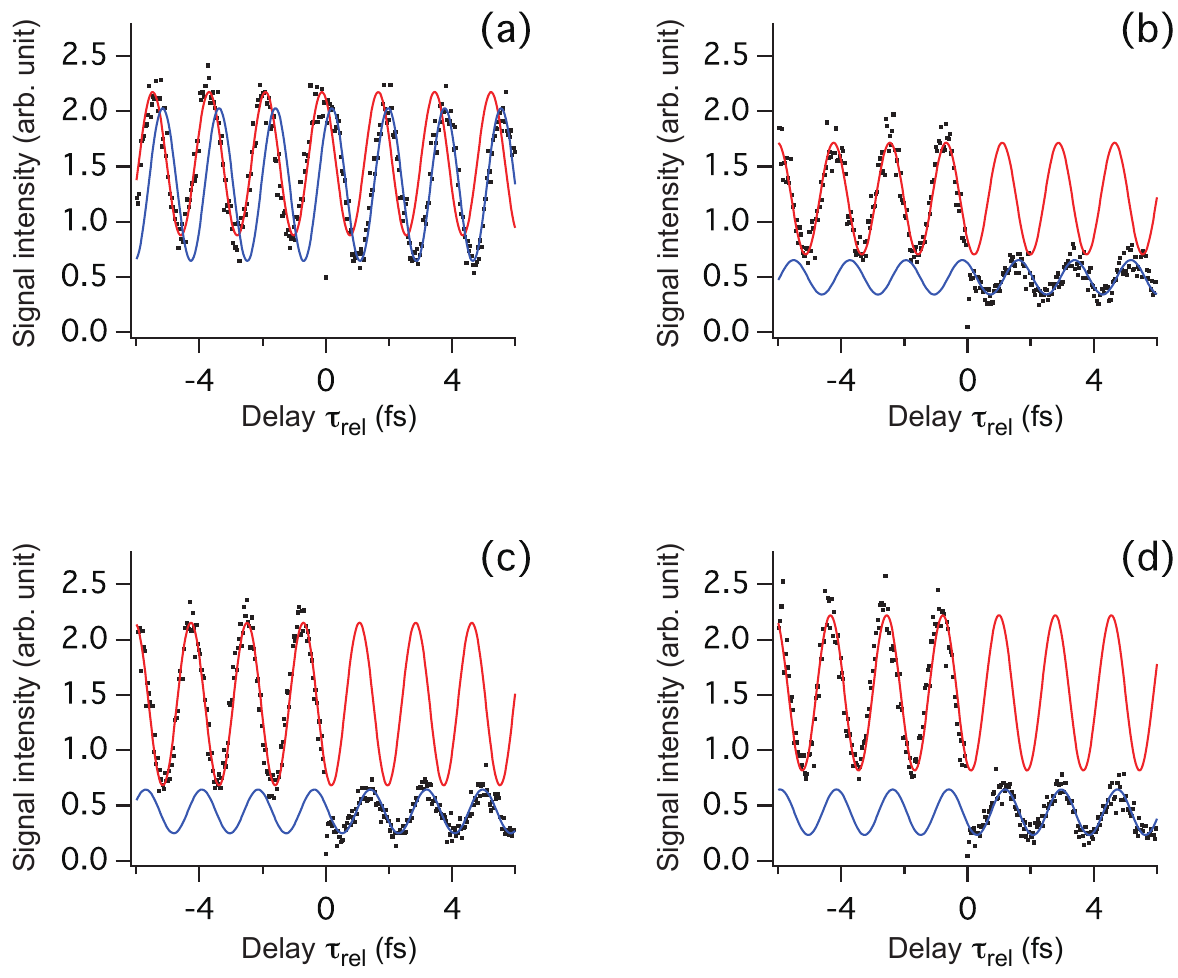}
\caption{Eigenstate interferograms not shown in Fig. 2.
(a) (Dotted line) Eigenstate interferograms of $v_{B}$=30 ($\tau_{\mathrm{rel}}<$0) and $v_{B}$=29 ($\tau_{\mathrm{rel}}>$0) with the NIR pulse shined at $\tau_{\mathrm{NIR}}$ $\sim$ 5.36 ps.
(Red line) Fitted sine curve for $v_{B}$=30. (Blue line) Fitted sine curve for $v_{B}$=29. 
(b) (Dotted line) Similar to (a) for $v_{B}$=30 ($\tau_{\mathrm{rel}}<$0) and $v_{B}$=31 ($\tau_{\mathrm{rel}}>$0) with the NIR pulse shined at $\tau_{\mathrm{NIR}}$ $\sim$ 5.07 ps. (Red line) Fitted sine curve for $v_{B}$=30. (Blue line) Fitted sine curve for $v_{B}$=31.
(c) (Dotted line) Similar to (b) without the NIR pulse. (Red line) Fitted sine curve for $v_{B}$=30. (Blue line) Fitted sine curve for $v_{B}$=31.
(d) (Dotted line) Similar to (b) with the NIR pulse shined at $\tau_{\mathrm{NIR}}$ $\sim$ 5.36 ps. (Red line) Fitted sine curve for $v_{B}$=30. (Blue line) Fitted sine curve for $v_{B}$=31.
The timing of the control pulse, $\tau_{\mathrm{control}}$, is given as $\tau_{0}$ + $\tau_{\mathrm{rel}}$  with $\tau_{0}$ set to $\sim$ 6.68 ps.
}
\vspace{10cm}
\end{figure}

\newpage

{\bf Video S1.}
Simulated wave-packet spreading suppressed by a train of strong laser pulses.
The wave packet here is an isolated three-level system composed of the eigenstates $v_{B}$ = 29, 30 and 31 of the iodine molecule with their population ratio 1:1:1 at the time origin $t$ = 0. The numerical details are provided in Supplementary Information. 
(Light blue) Free wave packet without NIR pulses. (Dark blue) Wave packet  controlled with a train of seven NIR pulses shined at $t$ =$(3n-1.5)T$ with $n$=1, 2, ..., 7, where $T$= 475 fs is a classical oscillation period of the wave packet.
Each NIR pulse has a Gaussian envelope with a temporal width of 0.05 $T$ (FWHM). 
The spreading of the wave packet is stopped almost completely by the NIR pulses.